\documentclass[twocolumn,preprintnumbers,prd,floatfix,nofootinbib]{revtex4}
\pdfoutput=1
\usepackage{hyperref}
\usepackage{graphicx}

\begin{document}

\title{Quantum Interference Effects Among Helicities at LEP-II and Tevatron}
\author{Matthew R. Buckley$^{1,2,3}$, Beate Heinemann$^{2,3}$, William Klemm$^{1,2,3}$, and Hitoshi Murayama$^{1,2,3}$} 
\affiliation{$^1$Institute for the Physics and Mathematics of the
  Universe, University of Tokyo, Kashiwa, Chiba 277-8568, Japan}
\affiliation{$^2$Department of Physics, University of California,
  Berkeley, CA 94720, USA} 
\affiliation{$^3$Physics Division, Lawrence Berkeley National
  Laboratory, Berkeley, CA 94720, USA}
\date{\today}

\begin{abstract}
  A completely model-independent method
  of obtaining information on the spin using the quantum interference
  effect among various helicity states was proposed in a recent paper. Here we point out
  that this effect should be demonstrable in the existing data on $e^-
  e^+ \rightarrow W^+ W^-$ at LEP-II and $p \bar{p} \rightarrow Z^0 + j$
  at Tevatron.
\end{abstract}
\preprint{IPMU 08-0017}
\preprint{UCB-PTH-08/07}
\maketitle

There are many reasons to expect that new particle degrees of freedom
will be discovered at the TeV energy scale (Terascale), starting with
the Large Hadron Collider (LHC) coming online later this year.  The
fact that the Terascale must have interesting physics has been known since Fermi's 1933 theory of nuclear beta decay which introduced a dimensionful constant $G_F \approx (0.3~\mbox{TeV})^{-1}$.
In its more modern incarnation, this constant represents the size of
the Bose--Einstein condensate that makes the universe a gigantic
superconductor. The analog of the Meissner effect then makes the range
of the weak force as short as a billionth of a nanometer.  

At the least we expect the gap excitation of the superconductor, the Higgs
boson, to be discovered at the LHC.  In addition, the quantum instability of this
energy scale suggests new particles below a TeV in order to protect it from
diverging to infinity.  Many theoretical frameworks have been proposed in the literature: new
strongly coupled gauge theory (technicolor \cite{Weinberg:1979bn,Susskind:1979}), fermionic dimensions of spacetime (supersymmetry \cite{Wess:1974tw}), bosonic dimensions of spacetime (extra dimensions \cite{ArkaniHamed:1998nn,Randall:1999ee}), new hidden extra symmetries (little Higgs \cite{ArkaniHamed:2001nc}), Higgless theories \cite{Csaki:2003dt,Csaki:2003zu} {\it etc}.  Many of
these also provide candidates for the mysterious dark matter of the
universe.  With great anticipation the community awaits the imminent
discovery of such exotic new particles in the upcoming LHC experiments.

Once new particles are discovered, determining what theoretical
framework they belong to is of foremost importance.  For this purpose
truly basic measurements will be required: mass, parity, and spin of the
new particles.  Among these, the spin measurement is both the key and the
most challenging.  Numerous studies exist that try to formulate
strategies for spin measurements at the LHC \cite{Barr:2004ze,Battaglia:2005zf,Smillie:2005ar,Battaglia:2005ma,Barr:2005dz,Wang:2006hk,Choi:2006mr,Alves:2007xt}.  Unfortunately, it is very difficult to avoid model-dependent assumptions in the proposed
measurement strategies.

In a recent paper \cite{Buckley:2007th}, three of us (M.B., W.K., H.M.) proposed a completely model-independent way of
obtaining information about spin at collider experiments.\footnote{This possibility was originally suggested in \cite{MurayamaTalk}.}  The key
element is quantum interference among various helicity
states of the new particle, which, to our surprise, has not been
discussed in the modern literature (see, however, \cite{jackson}).  We discussed how this method may
work to discriminate the smuon in supersymmetry or the Kaluza--Klein muon in
extra dimensions at the proposed International Linear Collider (ILC).

In this letter, we point out that the effectiveness of our proposed
method should be demonstrable in the existing data.  In particular,
$e^- e^+ \rightarrow W^+ W^-$ at LEP-II and $p \bar{p} \rightarrow
Z^0+j$ at Tevatron should allow highly significant studies of the
quantum interference among helicities, and demonstrate the spin-one
nature of the $W$ and $Z$ bosons without any model assumptions. As discussed in \cite{Buckley:2007th}, this method works particularly well close to the production threshold.  This is good news for the LHC, 
as new physics there will likely be dominated by the energy range just above threshold.

The proposed strategy is extremely simple.  In order to obtain
model-independent information about spin, or angular momentum in
general, we resort to the general principles of quantum mechanics.
The angular momentum operators generate spatial rotations; the
unitary operator $U(\vec{\phi}) =
e^{i\vec{J}\cdot\vec{\phi}/\hbar}$ rotates space around the axis
$\vec{\phi}$ by the angle $|\vec{\phi}|$.  If we choose the
rotation axis to be the momentum vector of a free particle, it
isolates the spin component because the orbital angular momentum is
always orthogonal to the momentum vector $\vec{L} \cdot \vec{p} =
(\vec{x} \times \vec{p}) \cdot \vec{p} = 0$.  Therefore, the angular
momentum along the momentum vector is nothing but its helicity, $h =
(\vec{s} \cdot \vec{p})/|\vec{p}|$.  The rotation around the momentum
axis by an angle $\phi$ therefore gives the phase $e^{ih\phi}$ to the
quantum mechanical amplitudes.

Obviously a single phase factor does not lead to a physical observable
since the probability does not depend on phases.  However, an
interference effect may pick up the {\it differences}\/ in phases among
interfering amplitudes.  Fortunately, particles produced in collisions
are often in a linear superposition of various helicity states,
which interfere when they decay into a common final state. This interference of different helicity states produces a cross section dependent on the coherent sum of individual matrix elements squared:
\begin{eqnarray}
&& \sigma \propto   \left|\sum_h {\cal M}_{\rm prod.}(h)  {\cal M}_{\rm decay}(h,\phi)\right|^2 \label{eq:sigmaphi} \\
&& {\cal M}_{\rm decay}(h,\phi)  =  e^{ih\phi} {\cal M}_{\rm decay}(h,\phi=0). \nonumber
\end{eqnarray}
Here ${\cal M}_{\rm prod.}(h)$ and ${\cal M}_{\rm decay}(h,\phi=0)$ are the production and decay matrix elements, which depend in detail on the helicity state $h$. However, all $\phi$ dependence has been factored out into the exponential. It is clear from this sum that the azimuthal angular dependance of the event distributions $N = \sigma \times {\cal L}$ (where $\cal L$ is the luminosity) is 
\begin{equation}
{dN\over{d\phi}} = {d\sigma\over {d\phi}} \times {\cal L} = A_0 +A_1\cos\phi + \cdots +A_n \cos(n \phi), \label{eq:sigmaexpan}
\end{equation}
where $n = \Delta h$ is the difference between the highest and lowest helicity states contributing to the sum in Eq.~(\ref{eq:sigmaphi}). In this way, we obtain an unambiguous lower limit on the
spin of the particle, $s \geq (\Delta h)/2$. As we will see, this limit is saturated, $s = \Delta h/2$, in the examples below, and the presence of the highest mode is clearly visible in collider data given sufficient statistics.

In the cases of $e^-e^+\to W^+W^-$ with leptons plus jets final states and $p\bar{p} \to Z^0 + j$ with decays to electrons, spin-$1$ particles are produced in a superposition of helicity states.\footnote{It is for this reason we cannot consider $p\bar{p}\to Z$ without jets. In such events, the $Z$ is produced in only one spin state, depending on the spin of the initial state quarks. While the cross section would contain a sum over $Z$ helicity, the sum would be incoherent.} In both cases, the event is fully reconstructable using the visible momentum in the event, and hence the angle $\phi$ can be fully determined from data.

\begin{figure}[t]
\centering
\includegraphics[width=\columnwidth]{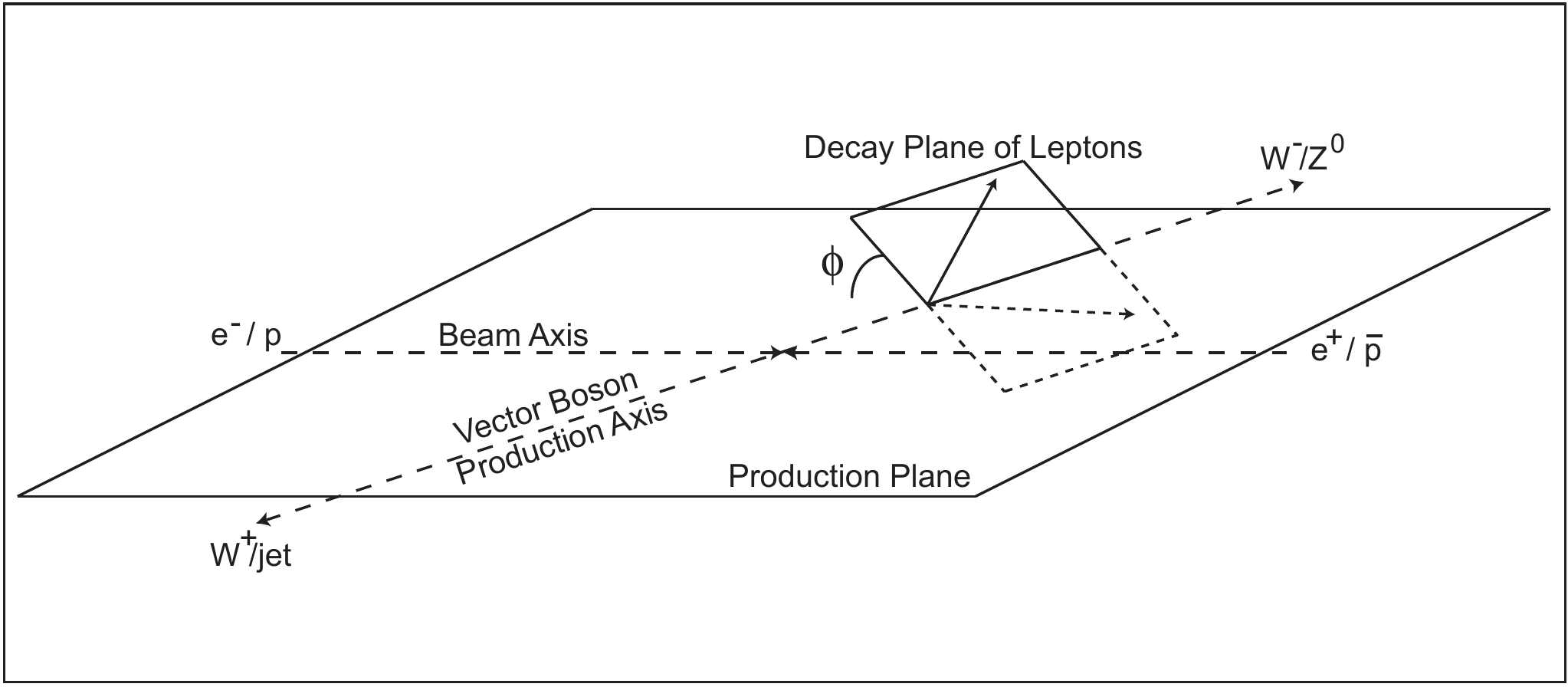}
\caption{The event kinematics of $e^-e^+ \to W^+W^- \to q\bar{q} \ell^\pm \nu$ at LEP-II and $p\bar{p} \to Z^0 +j \to e^-e^+ +j$ at Tevatron. The plane of pair produced vector bosons and the plane formed by the leptonic decay of one boson are shown. The angle $\phi$ is the relative azimuthal angle between these two planes, defined in the lab frame of the event, as defined in Eq.~(\ref{eq:phidef}). Positive $\phi$ are in the direction of the $e^-$ ($p$) beam momentum crossed with the $W^-$ ($Z^0$) momentum for LEP-II (Tevatron).} \label{fig:kinematics}
\end{figure}

The angle $\phi$ is defined in the lab frame of the event as the angle between the production plane described by the $W^+W^-$ or $Z^0 + j$ and the decay plane containing the leptonic decay products from the vector bosons. We define the positive $z$ axis in the lab frame of LEP-II (Tevatron) as the direction of $e^-$ (proton) beam, then the cosine of $\phi$ at LEP-II can be calculated as follows:
\begin{eqnarray}
\hat{n}_{\rm prod.} \equiv  \frac{\hat{z} \times \vec{p}_{W^\pm}}{|\hat{z} \times \vec{p}_{W^\pm}|}, & & \hat{n}_{\rm decay} \equiv \frac{\vec{p}_{W^\pm} \times \vec{p}_{\ell^\pm}}{|\vec{p}_{W^\pm} \times \vec{p}_{\ell^\pm}|} \nonumber \\
\cos \phi & = & \hat{n}_{\rm prod.}\cdot \hat{n}_{\rm decay} \label{eq:phidef},
\end{eqnarray}
where $\vec{p}_{\ell^\pm}$ is the charged lepton from the decay of the $W^\pm$ boson. The definition of $\phi$ at Tevatron is the same as in Eq.~(\ref{eq:phidef}) with the substitution of $Z^0$ for $W^\pm$. An arbitrary (but consistent) choice must be made to define which side of the production plane will contain positive $\phi$. For LEP-II, we chose this positive direction to be in the direction of $\hat{z}$ crossed with the momentum of the leptonically decaying $W^\pm$. Similarly, we chose the direction of the proton beam crossed with that of the $Z^0$ at Tevatron (see Fig.~\ref{fig:kinematics}). Based on our argument above, we expect to see cross sections for these events as in Eq.~(\ref{eq:sigmaexpan}) with $n=2$.\footnote{It should be noted that if the collider beams are identical, this choice of positive $\phi$ suffers from an ambiguity which maps $\phi \to \phi+\pi$. This may, for example, introduce difficulties in measuring $A_n$ ($n$ odd) parameters at LHC.}

\begin{table}[t]
\centering

\begin{tabular}{|c|r@{$\pm$}l|}
\hline
$\sqrt{s}$ (GeV) & \multicolumn{2}{c|}{${\cal L}$ (pb$^{-1}$)} \\ \hline
$182.25$ & $56.8 $ & $ 0.3$ \\ \hline
$188.63$ & $174.2 $ & $ 0.8$ \\ \hline
$191.58$ & $28.9$ & $ 0.1$ \\ \hline
$195.52$ & $79.9 $ & $ 0.4$ \\ \hline
$199.52$ & $86.3 $ & $ 0.4$ \\ \hline
$201.62$ & $41.9 $ & $ 0.2$ \\ \hline
$204.86$ & $81.4 $ & $ 0.4$ \\ \hline
$206.53$ & $133.2 $ & $ 0.6$ \\ \hline
\end{tabular}
\caption{LEP-II integrated luminosity $\cal L$ as a function of beam energy $\sqrt{s}$ \cite{Heister:2004wr}.} 
\label{tab:LEPlum}
\end{table}

The LEP-II luminosity from the years 1997-2000 \cite{Heister:2004wr} are reported in Table~\ref{tab:LEPlum}. The OPAL collaboration has observed 1574 events identified as $q\bar{q}e\nu$ and an additional 1573 $q\bar{q}\mu \nu$ events \cite{Abbiendi:2005eq}. Due to the low purity of the $q\bar{q}\tau \nu$ sample, we ignore those events. Similar data sets are available from the ALEPH \cite{Heister:2004wr}, DELPHI \cite{Abdallah:2003zm}, and L3 \cite{Achard:2004zw} collaborations.

The CDF collaboration has data for $Z^0+j$ consisting of 6203 events \cite{:2007cp} after selection cuts from $1.7\mbox{ fb}^{-1}$ of luminosity at $1.96~$TeV beam energy. D\O\ has a similar data set available  \cite{Abazov:2006gs}. A total luminosity of $8~\mbox{fb}^{-1}$ is expected to be available from Tevatron at the conclusion of data collection.

Parton level matrix elements for $W^+W^-$ and $Z^0+j$ (where the jet consists of a gluon or first generation (anti) quark at the parton level) production were calculated in HELAS \cite{Murayama:1992gi}, while the numerical integration program BASES \cite{Kawabata:1985yt} was used to determine the differential cross section and integrate over all other kinematic variables. For the simulation of the Tevatron results, a $K$ factor of $1.4$ was used to correct for higher order QCD effects, in accordance with \cite{:2007cp}, and CTEQ5L PDFs were implemented using LHAPDF \cite{Whalley:2005nh}. The Tevatron results and fits were confirmed using ALPGEN \cite{Mangano:2002ea}.

\begin{figure}[ht]
\centering

\includegraphics[width=0.5\columnwidth]{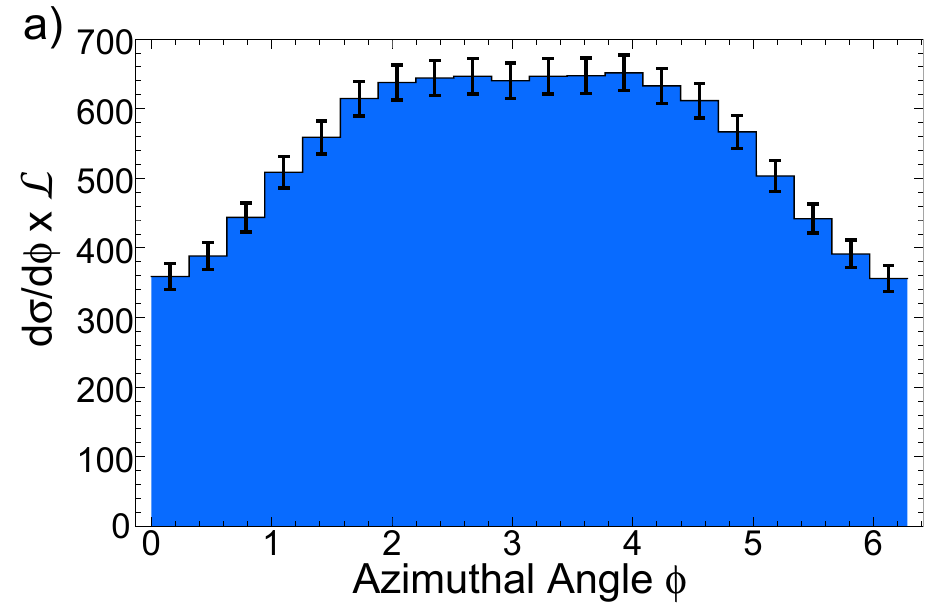}\includegraphics[width=0.5\columnwidth]{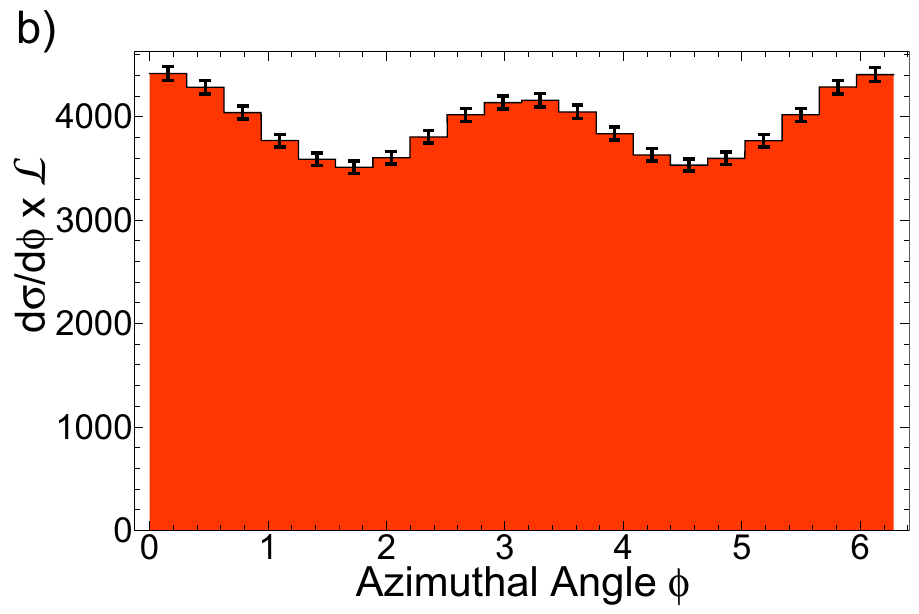}

\caption{Differential distribution of events $d\sigma/d\phi \times {\cal L}$ for a) $e^-e^+ \to W^+W^-\to q\bar{q} \ell^\pm \nu$ using the LEP-II run data in Table~\ref{tab:LEPlum}  and b) $p\bar{p} \to Z^0+j \to e^-e^+ +j$ with luminosity ${\cal L} = 1.7~\mbox{fb}^{-1}$. No cuts are applied on the LEP-II simulation, Tevatron results have $p_T > 30~\mbox{GeV}$ and $|\eta| < 2.1$ on the jet.} \label{fig:rawdist}
\end{figure}
\begin{table}[h]
\centering
\begin{tabular}{|c|r@{$\pm$}l|r@{$\pm$}l|}
\hline
 & \multicolumn{2}{c|}{LEP-II}& \multicolumn{2}{c|}{Tevatron} \\ \hline
$A_1/A_0$ & $-0.267$ & $0.023$ & ~$0.036$ & $0.009$ \\ \hline
$A_2/A_0$ & $-0.085$ & $0.025$ & $0.100$ & $0.009$ \\ \hline
$A_3/A_0$ & $0.000$ & $0.025$ & $0.000$ & $0.009$ \\ \hline
$A_4/A_0$ & $0.000$ & $0.026$ & $0.000$ & $0.010$ \\ \hline
\end{tabular}
\caption{Fits to the parameters $A_n$ in Eq.~(\ref{eq:sigmaexpan}) for the differential distributions of $e^-e^+ \to W^+W^-\to q\bar{q} \ell^\pm \nu$ (LEP-II) using the integrated luminosity in Table~\ref{tab:LEPlum}, and $p\bar{p} \to Z^0+j \to e^-e^+ +j$ (Tevatron) using ${\cal L} = 1.7~\mbox{fb}^{-1}$. Errors for each parameter are obtained by marginalizing over the other four parameters in the fit. No cuts are applied on the LEP-II simulation, Tevatron results have $p_T > 30~\mbox{GeV}$ and $|\eta| < 2.1$ on the jet.} \label{tab:rawfit}
\end{table}

The generated histograms are assigned Gaussian statistical error bars based on the realistic experimental luminosities. However, no statistical fluctuations are assigned to the central values. As a consequence, the fit results correspond to an average experiment \cite{deGouvea:1999wg}.

Before the application of cuts, the differential cross sections for the two processes of interest show a clear $\cos \phi$ and $\cos 2\phi$ dependence, as expected for the decays of spin-1 bosons. These distributions are shown in Fig.~\ref{fig:rawdist}. We then fit the parameters $A_0,A_1,A_2,A_3,$ and  $A_4$ in Eq.~(\ref{eq:sigmaexpan}) to the event distributions.\footnote{These fits are to the numerically integrated differential cross-section, not generated events.} For each of the five parameters $A_n$, $1$-$\sigma$ error bars are calculated after marginalizing over the other four. Results for the LEP-II and Tevatron simulations are shown in Table~\ref{tab:rawfit}; in order to compare simulations with different numbers of events, values of $A_n/A_0$ are reported rather than $A_n$. It it clear at this stage that the results are consistent with the decay of spin-1 bosons.


\begin{table}[t]
\centering

\begin{tabular}{|c|c|}
\hline
Jet transverse momentum & $p_{T,j} > 30~\mbox{GeV}$ \\ \hline
Jet $\eta$ & $|\eta| < 2.1$ \\ \hline
Invariant mass of lepton pair & $66 < m_{\ell \ell} < 116$~GeV \\ \hline
Central electron $\eta$ & $|\eta| < 1$ \\ \hline
Second electron $\eta$ & $|\eta| < 1$ or $1.2 < |\eta| <2.8$ \\ \hline
Electron $E_T$ & $E_T > 25 ~\mbox{GeV}$ \\ \hline
Electron isolation cuts & $\Delta R_{ej} > 0.7$ \\ \hline
\end{tabular}

\caption{Event selection cuts imposed by the CDF collaboration on $p\bar{p} \to Z^0+j \to e^-e^+ +j$ events. In each event, one electron must be central, and pass stricter cuts than the second electron. The isolation cut parameter is defined as $\sqrt{(\Delta \phi)^2 + (\Delta \eta)^2} \equiv \Delta R$ \cite{:2007cp}.}\label{tab:CDFcuts}
\end{table}

\begin{table}[t]
\centering

\begin{tabular}{|c|c|}
\hline
Lepton momentum & $p_\ell > 25~ \mbox{GeV}$ \\ \hline
Polar angle $\theta$ of final state particles & $|\cos\theta | <0.95$ \\ \hline
Neutrino energy fraction & $R_\nu > 0.07$ \\ \hline
Visible energy fraction & $R_{\rm vis} > 0.3$ \\ \hline
Neutrino transverse momentum & $p_{T,\nu} > 16~\mbox{GeV}$ \\ \hline
Lepton isolation & $\Delta R > 0.75,0.5,0.2$ \\ \hline
\end{tabular}

\caption{Event selection cuts imposed by the OPAL collaboration on $e^-e^+ \to W^+W^-\to q\bar{q} \ell^\pm \nu$ events. Energy fraction is defined as $R_\alpha  \equiv E_\alpha /\sqrt{s}$, where $\alpha$ is either the neutrino $\nu$ or the total visible energy. The lepton isolation cut was implemented using $\sqrt{(\Delta \phi)^2 + (\Delta \eta)^2} \equiv \Delta R$ with a range of $\Delta R$ values rather than limiting total energy deposited in cone surrounding the lepton as in \cite{Ackerstaff:1996nk}.}\label{tab:OPALcuts}
\end{table}

However, cuts must be applied to the events recorded at LEP-II and Tevatron, both due to detector geometry and in order to reduce background. These cuts will affect the azimuthal distribution present in $d\sigma / d\phi \times {\cal L}$, and so can obscure the signal necessary for spin measurements. 
The Tevatron cuts (Table~\ref{tab:CDFcuts}) were taken from the CDF experiment \cite{:2007cp}, while the OPAL \cite{Ackerstaff:1996nk} cuts (Table~\ref{tab:OPALcuts}) were used to simulate the LEP-II data. 

\begin{figure}[t]
\centering

\includegraphics[width=0.5\columnwidth]{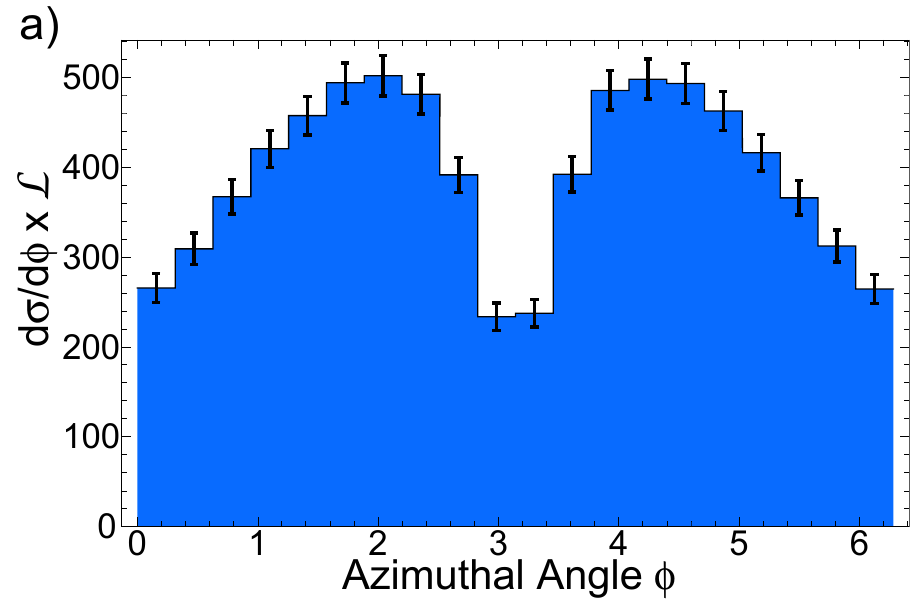}\includegraphics[width=0.5\columnwidth]{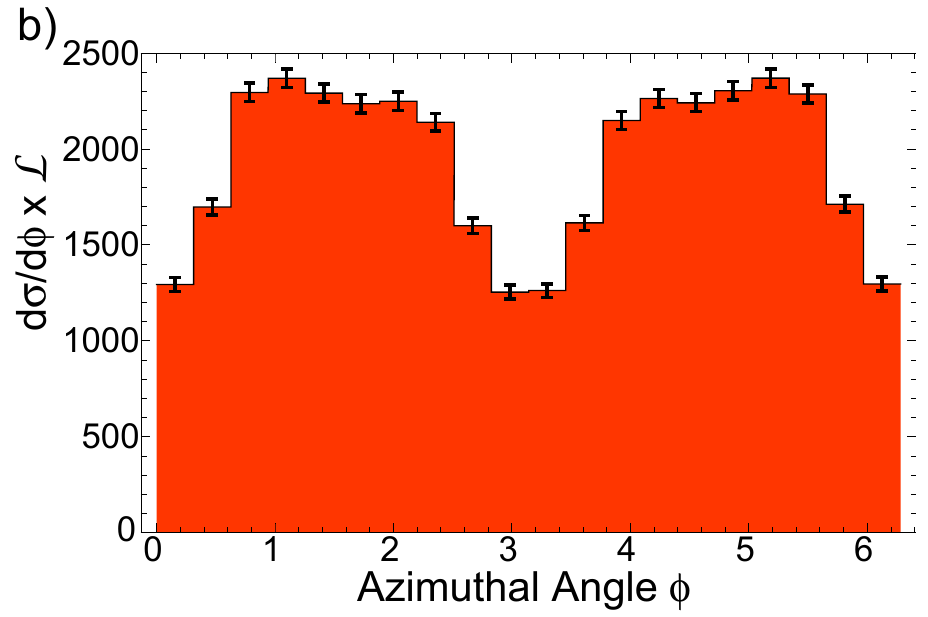}

\caption{Differential distributions for a) $e^-e^+ \to W^+W^-\to q\bar{q} \ell^\pm \nu$ with the cuts in Table~\ref{tab:OPALcuts} and $\Delta R = 0.75$ and b) $p\bar{p} \to Z^0+j \to e^-e^+ +j$ with the cuts from Table~\ref{tab:CDFcuts}. Luminosities are as in Fig.~\ref{fig:rawdist}.} \label{fig:cutsdist}
\end{figure}

Our simulation did not include parton showers or hadronization, so we could not implement the lepton isolation cut used by OPAL, which placed a limit on the total energy deposited in a cone centered on the lepton. Instead, we used a cut on $\Delta R \equiv \sqrt{(\Delta \phi)^2 + (\Delta \eta)^2}$ between the jet and the leptons. Three values for $\Delta R$ were used: $0.2$, $0.5$ and $0.75$, which gave total efficiencies for the cuts of $79\%$, $76\%$, and $72\%$ respectively. The cuts used by the OPAL collaboration had an efficiency of 85\% for final states with an electron and 89\% for muons. The distributions of the Tevatron and LEP-II (with $\Delta R = 0.75$) simulations after cuts are shown in Fig.~\ref{fig:cutsdist}.

Fitting the distributions to Eq.~(\ref{eq:sigmaexpan}), we find the results in Table~\ref{tab:cutfit}.  These results clearly show that the imposed cuts introduce spurious high frequency modes. The corresponding non-zero $A_3$ and $A_4$ components may naively be confused for evidence of spin-2 particles. However, the cuts are responsible for introducing new $\phi$ dependence by selecting out new directions relative to the production axis of the gauge bosons. 

\begin{figure}[t]
\centering

\includegraphics[width=\columnwidth]{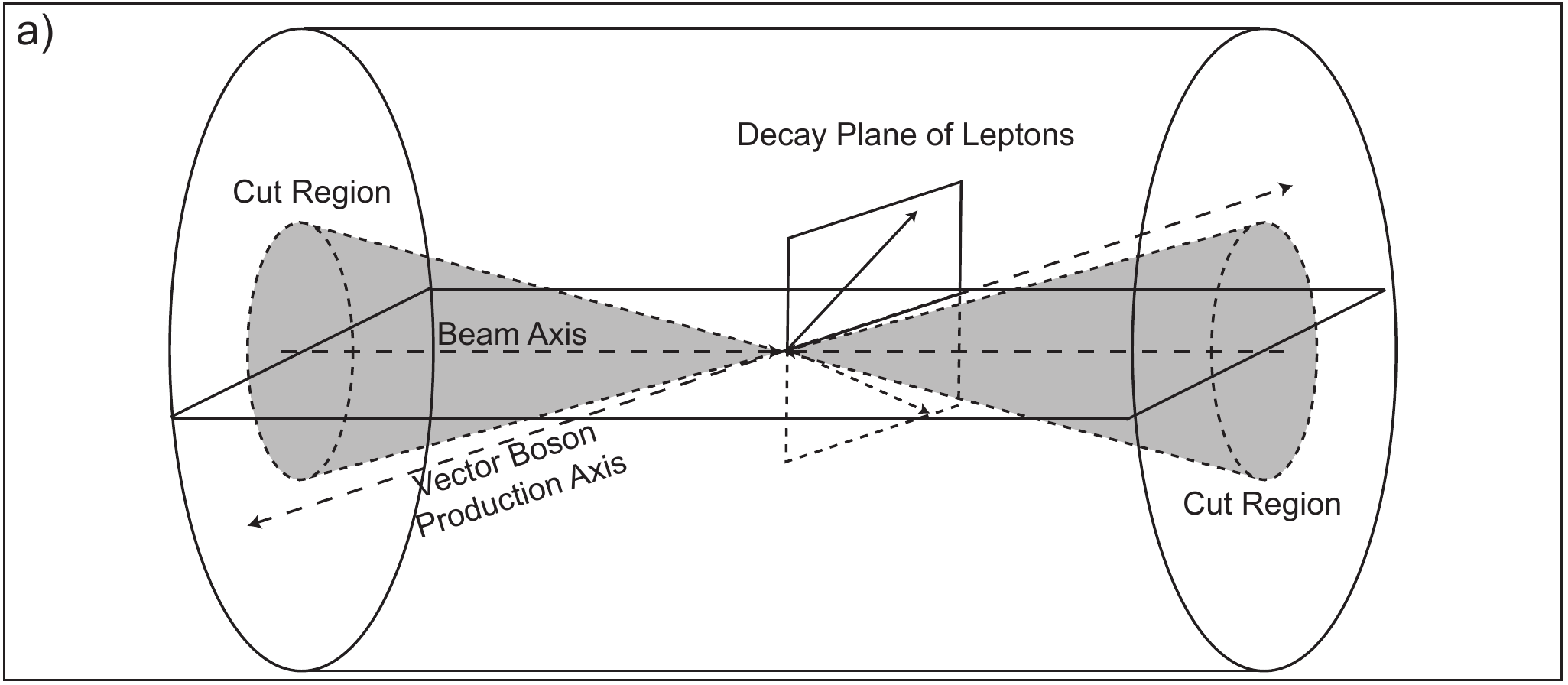}

\includegraphics[width=\columnwidth]{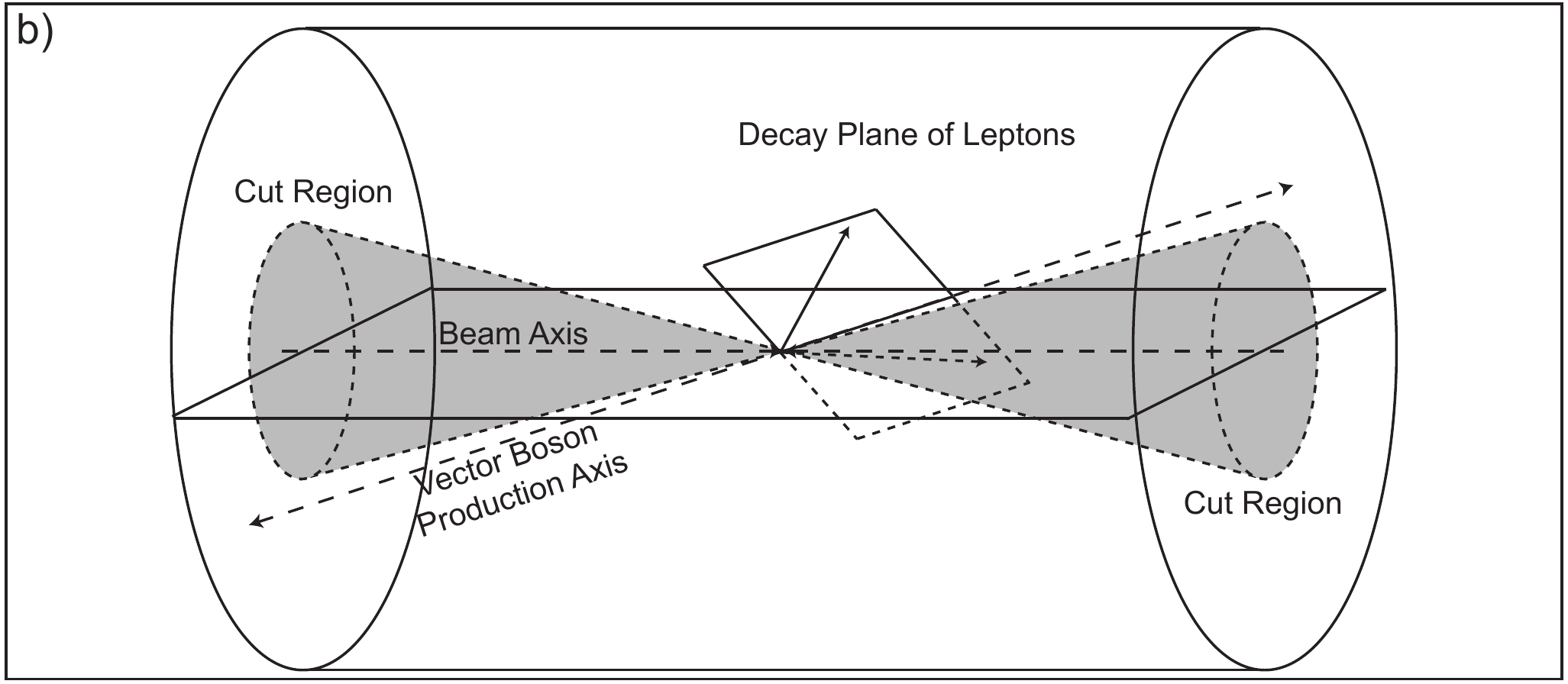}
\caption{A depiction of the detector volume demonstrating the rotational dependence induced by the cuts. The shaded forward regions (large values of $|\eta|$ and $|\cos\theta|$) are inaccessible due to detector geometry and background cuts. Two sample events are depicted in a) and b). These events are kinematically identical in the boson rest frame save for a rotation in $\phi$. The event a) survives the cuts, while the event b) fails.}\label{fig:cuts}
\end{figure}

We illustrate this effect for the case of cuts in the forward direction (large $|\eta|$ and $|\cos \theta|$) in Fig.~\ref{fig:cuts}. Here we see two decays which are kinematically identical in the boson rest frame save for azimuthal rotations. In Fig.~\ref{fig:cuts}a, the event survives the cuts, as neither lepton lies sufficiently close to the $z$ axis. However, in Fig.~\ref{fig:cuts}b, rotating the decay plane about the axis of the boson momentum yields an event which is eliminated by the cuts. This is the source of unwanted $\phi$ dependences in the differential distributions with cuts. Similar problems arise due to isolation cuts, which depend on the proximity of the leptons to the other particles in the final state, as well as cuts on leptonic transverse momentum.

Since this $\phi$ dependence did not arise from the quantum interference of helicity amplitudes, we cannot expect the $\phi$ dependence of the cross section to accurately reflect the spin of the decaying particles. Thus non-zero $A_3$ and $A_4$ components do not indicate a higher spin state, but rather a breakdown of the proposed spin-measurement technique.

The solution to this problem is relatively straightforward. For new azimuthal dependences to be avoided, the cuts cannot pick out `special' directions relative to the original momentum of the decaying boson. Therefore we impose `rotationally invariant cuts' in which we require that each event not only passes the experimental cuts but continues to do so when the decay plane is rotated around the boson production axis.
\begin{table}[t]
\centering
\begin{tabular}{|c|r@{$\pm$}l|r@{$\pm$}l|r@{$\pm$}l|}
\hline
 & \multicolumn{6}{c|}{LEP-II} \\ \hline
 & \multicolumn{2}{c|}{$\Delta R =0.75$}  & \multicolumn{2}{c|}{$\Delta R =0.5$} & \multicolumn{2}{c|}{$\Delta R =0.2$} \\ \hline
$A_1/A_0$ & $-0.082$ & $0.025$ & $-0.082$ & $0.026$ & $-0.082$ & $0.025$  \\ \hline
$A_2/A_0$ & $-0.293$ & $0.026$ & $-0.302$ & $0.027$ & $-0.308$ & $0.026$  \\ \hline
$A_3/A_0$ & $0.110$ & $0.027$ & $0.114$ & $0.028$ & $0.117$  & $0.028$  \\ \hline
$A_4/A_0$ & $-0.099$ & $0.028$ & $-0.099$ & $0.029$ & $-0.096$ & $0.029$ \\ \hline
\end{tabular}
\begin{tabular}{|c|r@{$\pm$}l|}
\hline
  & \multicolumn{2}{c|}{Tevatron} \\ \hline
$A_1/A_0$  & $0.029$ &  $0.012$ \\ \hline
$A_2/A_0$ & $-0.277$ & $0.012$ \\ \hline
$A_3/A_0$  & $-0.021$ & $0.013$ \\ \hline
$A_4/A_0$ & $-0.123$ & $0.014$ \\ \hline
\end{tabular}
\caption{Fits of the differential distribution of $e^-e^+ \to W^+W^-\to q\bar{q} \ell^\pm \nu$ (LEP-II) with the cuts in Table~\ref{tab:OPALcuts} and $p\bar{p} \to Z^0+j \to \ell^-\ell^+ +j$ (Tevatron) with the cuts in Table~\ref{tab:CDFcuts} to parameters $A_n$ in Eq.~(\ref{eq:sigmaexpan}). Luminosities are as in Table~\ref{tab:rawfit}. $1$-$\sigma$ errors for each parameter are obtained by marginalizing over the other four parameters in the fit.} \label{tab:cutfit}
\end{table}
This avoids the introduction of a new directional dependence since we restrict ourselves to only those events which could never overlap the forbidden regions of the detector regardless of orientation. However, these cuts are very inefficient: the cuts on LEP-II data preserve only 12\% of the original events, while the cuts for the Tevatron leave less than 1\%.

The CDF cuts are very inefficient due to the small allowed $|\eta|$ region for the central electron (see Table~\ref{tab:CDFcuts}). Recent preliminary CDF measurements have demonstrated that the cuts can be relaxed while still maintaining a background level of less than $5\%$ \cite{CDF}. These loosened cuts are identical to those in Table~\ref{tab:CDFcuts} for $p_T$ and $\eta$ of the jet and the invariant mass of $m_{\ell\ell}$. However, the central lepton is allowed $E_T > 20$~GeV and $|\eta|<2.6$, while the second electron must have $E_T > 10$~GeV and $|\eta|<2.6$. If both leptons have $2.6>|\eta|>1.0$, $E_T$ must be greater than $25$~GeV. Finally, $\Delta R_{ej}$ must be greater than $0.4$. With these relaxed numbers, the total number of events in the simulated sample is 5821 and the efficiency of the rotationally invariant cuts is 18\%. 

\begin{figure}[t]
\centering

\includegraphics[width=0.5\columnwidth]{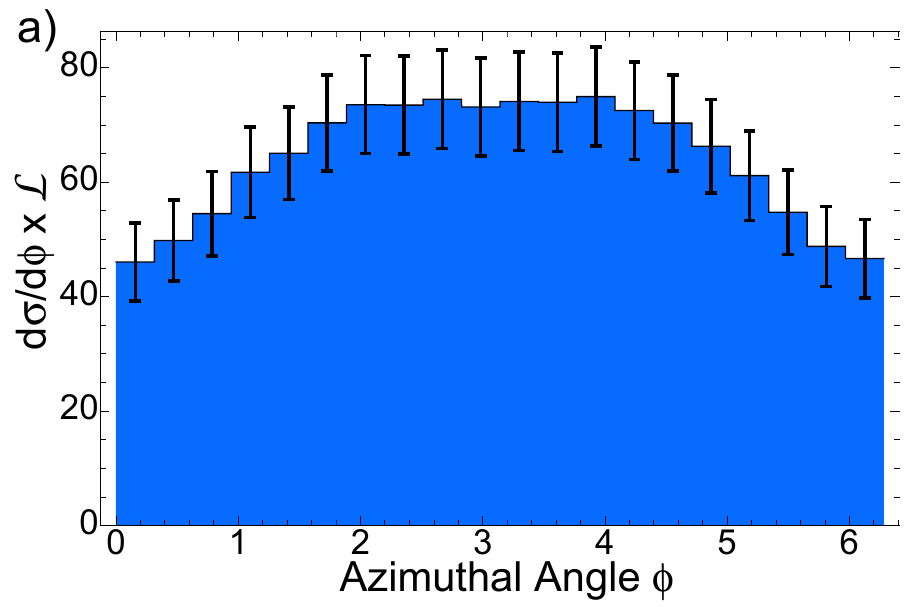}\includegraphics[width=0.5\columnwidth]{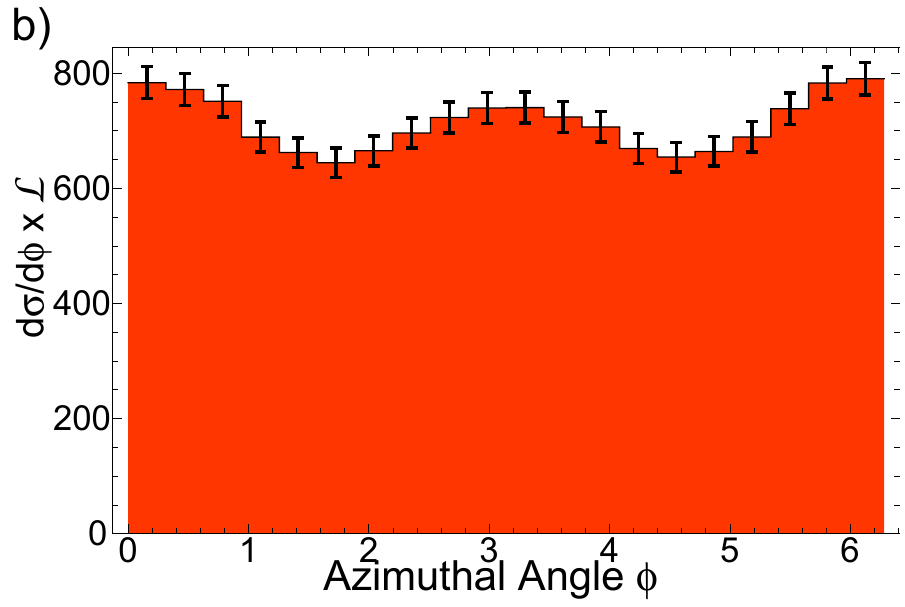}

\caption{Differential distributions for a) $e^-e^+ \to W^+W^-\to q\bar{q} \ell^\pm \nu$ and b) $p\bar{p} \to Z^0+j \to e^-e^+ +j$ requiring rotationally invariant cuts. Luminosities are as in Fig.~\ref{fig:rawdist}.} \label{fig:rotdist}
\end{figure}

The result of these rotationally invariant cuts on the LEP-II and Tevatron data are shown in Fig.~\ref{fig:rotdist} (compare to Fig.~\ref{fig:rawdist}). Table~\ref{tab:rotfit} confirms that this technique restores the $\phi$ dependence expected by the interference argument. 

In the case of the Tevatron results with loosened cuts, the data is clearly consistent with the $Z$ being a spin-1 vector boson. The $A_1$ parameter is non-zero at $1.8\sigma$, the $A_2$ parameter is non-zero at nearly $4\sigma$, and the higher modes are consistent with zero. It is important to recall that a lower bound on the spin is obtained from the {\it highest} non-zero mode, therefore the $4\sigma$ signal in $A_2$ is far more important then the $1.8\sigma$ deviation from zero in $A_1$.

\begin{table}[h]
\centering
\begin{tabular}{|c|r@{$\pm$}l|r@{$\pm$}l|r@{$\pm$}l|}
\hline
 & \multicolumn{6}{c|}{LEP-II} \\ \hline
 & \multicolumn{2}{c|}{$\Delta R =0.75$}  & \multicolumn{2}{c|}{$\Delta R =0.5$} & \multicolumn{2}{c|}{$\Delta R =0.2$} \\ \hline
$A_1/A_0$ & $-0.215$& $0.069$ & $-0.214$& $0.060$ & $-0.207$ & $0.053$  \\ \hline
$A_2/A_0$ & $-0.068$& $0.071$ & $-0.071$& $0.062$ & $-0.072$& $0.055$  \\ \hline
$A_3/A_0$ & $0.000$& $0.073$ & $0.000$& $0.064$ & $0.000$ & $0.057$  \\ \hline
$A_4/A_0$ & $0.000$ & $0.075$ & $0.000$& $0.065$ & $0.000$& $0.058$ \\ \hline
\end{tabular}

\begin{tabular}{|c|r@{$\pm$}l|}
\hline
  & \multicolumn{2}{c|}{Tevatron} \\ \hline
$A_1/A_0$  & ~$0.039$ & $0.022$ \\ \hline
$A_2/A_0$ & $0.083$ & $0.021$ \\ \hline
$A_3/A_0$  & $0.000 $ & $0.022$ \\ \hline
$A_4/A_0$ & $0.000 $ & $ 0.023$ \\ \hline
\end{tabular}
\caption{Fits of the differential distribution of $e^-e^+ \to W^+W^-\to q\bar{q} \ell^\pm \nu$ (LEP-II) and $p\bar{p} \to Z^0+j \to \ell^-\ell^+ +j$ (Tevatron) to the parameters $A_n$ in Eq.~(\ref{eq:sigmaexpan}), requiring events that pass the cuts in Tables~\ref{tab:OPALcuts} and \ref{tab:CDFcuts} (with relaxed $E_T$, $|\eta|$, and $\Delta R$ cuts as described in the text) after rotation about the momentum axis of the decaying vector boson. The luminosities are the same as in Tables~\ref{tab:rawfit} and \ref{tab:cutfit}. $1$-$\sigma$ errors for each parameter are obtained by marginalizing over the other four parameters in the fit.} \label{tab:rotfit}
\end{table}

From these results there is always the possibility that the parent $Z$ is a higher spin particle and that some conspiracy amongst the matrix elements in Eq.~(\ref{eq:sigmaphi}) prevents the $A_3$ and $A_4$ terms from appearing in the sum. In this interpretation, we can still state unambiguously that the $Z$ is {\it at least} spin-1, and that the data suggest it is not of higher spin. 

Higher statistics would allow a reduction of error bars and increase our confidence in the result correspondingly. Using, for example, the estimated total integrated luminosity of $8~\mbox{fb}^{-1}$ for the Tevatron, the parameters have the values shown in Table~\ref{tab:rotfitlum}. Another possibility is to use the muon decays of the $Z^0$. However, the rotationally invariant cuts will likely take a high toll on such events, as the muon tracking system at CDF extends only up to $|\eta| = 1.5$ \cite{Ginsburg:2004fa}.

The situation with the LEP-II simulation is more complicated. While the $A_1$ parameters are non-zero at over $3\sigma$, the $A_2$ parameters differ from zero by only one standard deviation. A larger data set would of course solve this problem. As all four LEP-II experiments (ALEPH, DELPHI, L3, and OPAL) have approximately equal statistics available, a two-fold increase in the statistical significance could be achieved by combining the events from these collaborations; the resulting ratios $A_n/A_0$ are shown in Table~\ref{tab:rotfitlum}.


Another possibility is that some reduction in required cuts would increase the efficiency of the rotationally invariant cuts without greatly degrading the sample purity. A likely candidate for this in our analysis is the $\Delta R$ cut, which was introduced as a stop-gap measure to approximate the jet-lepton proximity cut used in the OPAL analysis. However, even with the value of $\Delta R = 0.2$, the efficiency of the cut is lower than the $85\%$ reported by OPAL. Setting $\Delta R = 0$ is clearly an unrealistic cut, but as demonstrated in Table~\ref{tab:rotfitlum} indicates the possibilities offered by higher statistics.

In conclusion, we have demonstrated that the quantum interference among the matrix elements of different helicity states provides model-independent probe of particle spin. Using realistic data sets, rotationally invariant cuts can be implemented which correct for the spurious high-frequency noise introduced by the cuts imposed by detector geometry and background reduction. Though these techniques come at a price in terms of efficiency, it seems possible to relax the cuts in such a way that the weak gauge boson spins can be measured at sufficient significance at current colliders.

Measurements of the spin of new particles is expected to be a critical discriminator of new physics at the LHC. As a result, techniques such as the one proposed here are very important. Though the spins of the $W$ and $Z$ bosons are not in doubt, we find it encouraging that this new method can be tested on the available data. Such work would be of great use in the coming LHC era.

\begin{table}[h]
\centering
\begin{tabular}{|c|r@{$\pm$}l|r@{$\pm$}l|r@{$\pm$}l|}
\hline
 & \multicolumn{4}{c|}{LEP-II} &  \multicolumn{2}{c|}{Tevatron}  \\ \hline
 &  \multicolumn{2}{c|}{Combined}  &  \multicolumn{2}{c|}{$\Delta R = 0$} &  \multicolumn{2}{c|}{${\cal L} = 8~\mbox{fb}^{-1}$} \\ \hline
$A_1/A_0$  & $-0.207$ & $0.027$ & $-0.211 $ & $ 0.050$ & ~$0.039 $ & $ 0.010$ \\ \hline
$A_2/A_0$  & $-0.072$ & $ 0.028$ & $-0.081 $ & $ 0.052$ & $0.083 $ & $ 0.010$ \\ \hline
$A_3/A_0$  & $0.000 $ & $ 0.028$ & $0.000 $ & $ 0.053$ & $0.000 $ & $ 0.010$ \\ \hline
$A_4/A_0$  & $0.000 $ & $ 0.029$ & $0.000 $ & $ 0.054$ & $0.000 $ & $ 0.010$ \\ \hline
\end{tabular}
\caption{Fits of the differential distribution to the parameters $A_n$ in Eq.~(\ref{eq:sigmaexpan}) for: $e^-e^+ \to W^+W^-\to q\bar{q} \ell^\pm \nu$ with the jet-lepton cut parameter $\Delta R=0.2$ and combining the data sets of ALEPH, DELPHI, L3, and OPAL (LEP-II, Combined), $e^-e^+ \to W^+W^-\to q\bar{q} \ell^\pm \nu$ with the OPAL data set and $\Delta R$ set to zero  (LEP-II, $\Delta R =0$), and $p\bar{p} \to Z^0+j \to \ell^-\ell^+ +j$ with $8~\mbox{fb}^{-1}$ integrated luminosity (Tevatron).
We require all events pass the cuts in Tables~\ref{tab:OPALcuts} (with $\Delta R $ as indicated) and \ref{tab:CDFcuts} (with relaxed $E_T$ and $|\eta|$ cuts as described in the text) after rotation about the momentum axis of the decaying vector boson. $1$-$\sigma$ errors for each parameter are obtained by marginalizing over the other four parameters in the fit.} \label{tab:rotfitlum}
\end{table}

\begin{acknowledgments}
  This work was supported in part by World
 Premier International Research Center Initiative (WPI Program),
 MEXT, Japan, in part by the U.S. DOE under Contract
 DE-AC03-76SF00098, and in part by the NSF under grant PHY-04-57315.
\end{acknowledgments}

\end{document}